\documentclass[a4paper,11pt]{article}
\usepackage{latexsym}
\usepackage{amsmath}
\usepackage{amssymb}
\usepackage{graphicx}
\usepackage{wrapfig}
\begin{document}

\baselineskip=7mm


\bigskip

\centerline{\bf C and P violations in the PNJL model}

\centerline{H. Kouno$^1$, Y. Sakai$^2$, T. Sasaki$^2$, K. Kashiwa$^2$ and M. Yahiro$^2$}
\centerline{\it 1) Department of Physics, Saga University, Saga 840-8502, Japan}
\centerline{\it 2) Department of Physics, Kyushu University, Fukuoka 812-8581, Japan}
\baselineskip=3mm

~

\small

Using the PNJL model, we investigate C and P violations when $\theta =\pi$ and $\Theta =\mu/(iT)=\pi/3$, where $T$, $\theta$ and $\mu$ are the temperature, the parameter of $\theta$-vacuum and the quark number chemical potential, respectively. 
It is shown that the C violation and the P restoration happen almost simultaneously at $\theta=\pi$ and $\Theta =\pi/3$, if the deconfinement and the chiral symmetry restoration happen almost simultaneously at $\theta =\Theta =0$. 

~

~

To say exactly, in the realistic QCD, the deconfinement and the chiral restoration transitions have no exact symmetry which characterized these transitions, although the chiral symmetry is a good approximate symmetry. 
This fact makes it difficult to study these transitions analytically. 
In particular, the relation between these two transition is not clear. 
(See Ref. [1] and references therein. )

In pure gauge system, $\mathbb{Z}_3$ symmetry is the symmetry related to deconfinement transition and Polyakov-loop $\Phi$ is the order parameter of the deconfinement transition. 
In full QCD which includes dynamical quark, the system is not $\mathbb{Z}_3$-invariant any more. 
However, instead of the $\mathbb{Z}_3$ symmetry, there is the Roberge-Weiss (RW) periodicity [2] with respect to the imaginary chemical potential $\mu=i\Theta T$ where $T$ is the temperature of the system and $\Theta$ is a dimensionless real parameter. 
Roberge and Weiss found that QCD has a periodicity $\Omega_{\rm QCD}(\Theta)=\Omega_{\rm QCD}(\Theta+2\pi k/3)$, where $k$ is an arbitrary integer and $\Omega_{\rm QCD}$ is the thermodynamic potential in QCD. 

Roberge and Weiss also found that there is a phase transition from one $\mathbb{Z}_3$ -pseudo-degenerate vacuum to another at $\Theta =(2k+1)\pi/3$, if $T$ is larger than some critical temperature $T_{\rm RW}$ [2]. 
The results of Roberge and Weiss are based on the perturbative QCD (pQCD) and the strong coupling QCD. 
Afterward, the RW periodicity and the RW transition have been confirmed by the lattice QCD (LQCD) simulation. 
(See the references in [1]. ) 
Recently, in Ref. [3], it was pointed out that the deconfinement transition at zero chemical potential could be regarded as a remnant of the RW transition at $\Theta =\pi/3$, in which spontaneous breaking of the Charge conjugation (C) symmetry takes place. 
A $\Theta$-odd quantity such as a phase $\psi$ of modified Polyakov loop $\Psi$ which defined by $\Psi \equiv \Phi e^{i\Theta}$ is an order parameter of this phase transition.   

On the other hand, Dashen pointed out [4] that spontaneous parity (P) violation could arise at $\theta =\pi$, where $\theta$ is the parameter of $\theta$-vacuum.  
Using the NJL model, Boer and Boomsma analyzed the Dashen's phenomenon [5]. 
A parity odd quantity such as $\eta$ condensate is an order parameter of this phase transition. 
At low temperature, $\eta$ is nonzero and the P symmetry is broken. 
At high temperature, the $\eta$ condensate vanishes and the P symmetry is restored. 
This behavior of $\eta$ resembles the behavior of the chiral condensate $\sigma$ at $\theta=0$ which is large at low temperature and (almost) vanishes at low temperature. 
It seems that the chiral symmetry breaking $\theta =0$ is related to the P breaking at $\theta =\pi$. 
Therefore, studies on the relation between the C and P violations may give us an insight on the relation between the deconfinement and the chiral restoration at $\theta =\Theta =0$ and vice versa. 

In this study, using the PNJL model [6] and the EPNJL model [1], we investigate spontaneous C and P violations at $\Theta =\pi/3$ and $\theta =\pi$. 
We compare the result with the deconfinement and the chiral symmetry breaking at $\theta =\Theta =0$. 
Detail descriptions of the analyses will appear in the forthcoming paper [7]. 
The definitions of the parameters of the PNJL(EPNJL) model are the same as those in Ref. [1] and the values for the parameters are also the same as those in Ref. [1] if it is not mentioned explicitly. 
We put $c=0.2$ where the definition of the parameter $c$ is the same as that in ref. [5]. 

Figure 1(a) shows the $T$-dependence of chiral condensate $\sigma$ and Polyakov loop $\Phi$ at $\theta =\Theta =0$ calculated by using the original PNJL model. 
Since both of the deconfinement and the chiral restoration transitions are crossover, the transition temperatures are defined by the positions of the peaks of the susceptibilities of $\Phi$ and $\sigma$ [1]. 
The transition temperature $T_\chi$ of the chiral restoration is larger than the transition temperature $T_{\rm d}$ of the deconfinement. 
In this case, $\Delta T_{\rm \chi d}\equiv T_\chi -T_{\rm d}\sim 40$MeV. 

Figure 2(a) shows the $T$-dependence of $\eta$ condensate and a phase $\psi$ of modified Polyakov loop $\Psi$ at $\theta =\pi$ and $\Theta =\pi/3$ calculated by using the original PNJL model. 
At low temperature ($T<T_{\rm P}$), $\eta$ is finite and P symmetry is spontaneously broken, while, at high temperature ($T>T_{\rm C}$),  $\psi$ is finite and C symmetry is spontaneously broken. 
The transition temperature $T_{\rm P}$ of the P restoration is larger than the transition temperature $T_{\rm C}$ of the C violation. 
In this case, $\Delta T_{\rm PC}\equiv T_{\rm P}-T_{\rm C}\sim 80$MeV. 

Very recently, it is pointed out that $\Delta T_{\rm \chi d}$ becomes smaller if the coupling constant $G_{\rm s}$ of the four quark interaction in the PNJL model replaced by [1] 
$$ G_{\rm s}(\Phi )=G_s[1-\alpha_1\Phi\Phi^*-\alpha_2(\Phi^3+{\Phi^*}^3)], $$
where $G_{\rm s}$, $\alpha_1$ and $\alpha_2$ are constant parameters. 
Here, the new model is called entanglement PNJL (EPNJL) model. 
Figure 1(b) shows the $T$-dependence of chiral condensate $\sigma$ and Polyakov loop $\Phi$ at $\theta =\Theta =0$ calculated by using the EPNJL model. 
In this case, $\Delta T_{\rm \chi d}\equiv T_\chi -T_{\rm d}\sim 0$MeV. 

Figure 2(b) shows the $T$-dependence of $\eta$ condensate and a phase $\psi$ of modified Polyakov loop $\Psi$ at $\theta =\pi$ and $\Theta =\pi/3$ calculated by using the EPNJL model. 
In this case, $\Delta T_{\rm PC}\equiv T_{\rm P}-T_{\rm C}\sim 8$MeV. 
Comparing figures 1(b) and 2(b), we see that the C violation and the P restoration happen almost simultaneously at $\theta=\pi$ and $\Theta =\pi/3$, if the deconfinement and the chiral symmetry restoration happen almost simultaneously at $\theta =\Theta =0$. 
This result suggests that there may be a strong correlation between the C violation and the P restoration at $\theta =\pi$ and $\Theta =\pi/3$ if the deconfinement and the chiral restoration are strongly correlated at $\theta =\Theta =0$. 

~

\begin{flushleft}

\centerline{\bf References}

[1] Y. Sakai, T. Sasaki, H. Kouno and M. Yahiro, arXiv:1006.3648 [hep-ph] (2010). 

[2] A. Roberge and N. Weiss, Nucl. Phys. B {\bf 275}, 734 (1986). 

[3] H. Kouno, Y. Sakai, K. Kashiwa and M. Yahiro,  J. Phys. G: Nucl. Part. Phys. {\bf 36}, 115010(2009). 

[4] R. Dashen, Phys. Rev. D {\bf 3}, 1879 (1971). 

[5] D. Boer and K. Boomsma, Phys. Rev. D {\bf 78}, 054027(2008). 

[6] K. Fukushima, Phys. Lett. B {\bf591}, 277 (2004). 

[7] H. Kouno, Y. Sakai, T. Sasaki, K. Kashiwa and M. Yahiro,  in preparation.  

\end{flushleft} 


\begin{figure}
\begin{center}
 \includegraphics[width=6cm]{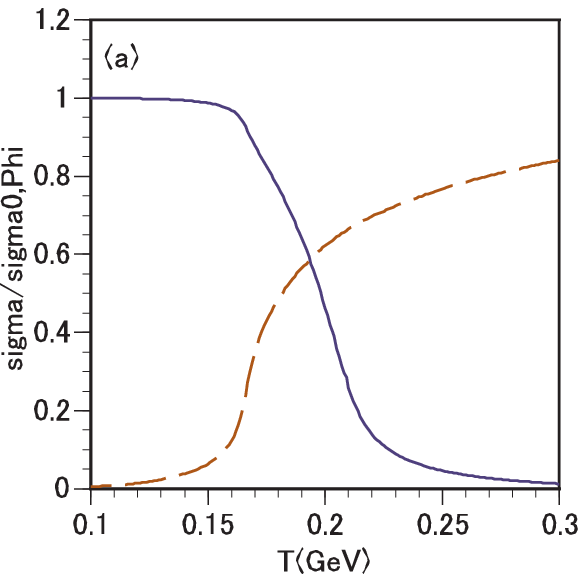} 
 \includegraphics[width=6cm]{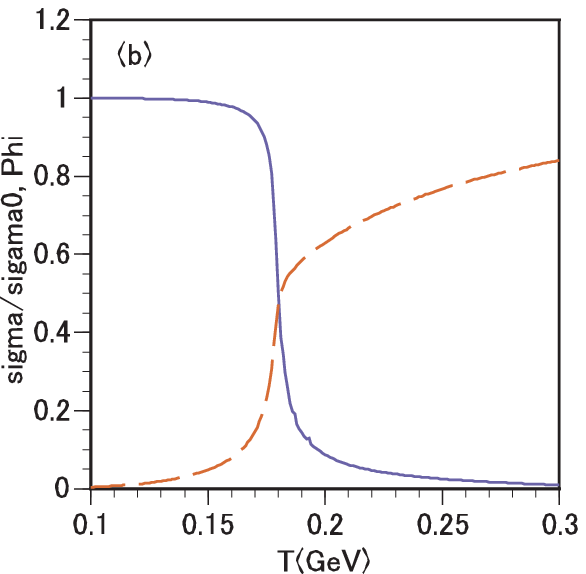} 
\end{center}
\caption{$T$-dependence of the chiral condensate $\sigma$ (solid curve) and the Polyakov loop $\Phi$ (dashed curve) at $\theta =\Theta =0$. 
$\sigma$ is normalized by $\sigma_0\equiv \sigma (T=0)$. 
(a) Original PNJL model with $T_0=200$MeV. (b) EPNJL model with $\alpha_1=\alpha_2=0.2$ and $T_0=200$MeV. 
}
\label{fig1}
\end{figure}
\begin{figure}
\begin{center}
 \includegraphics[width=6cm]{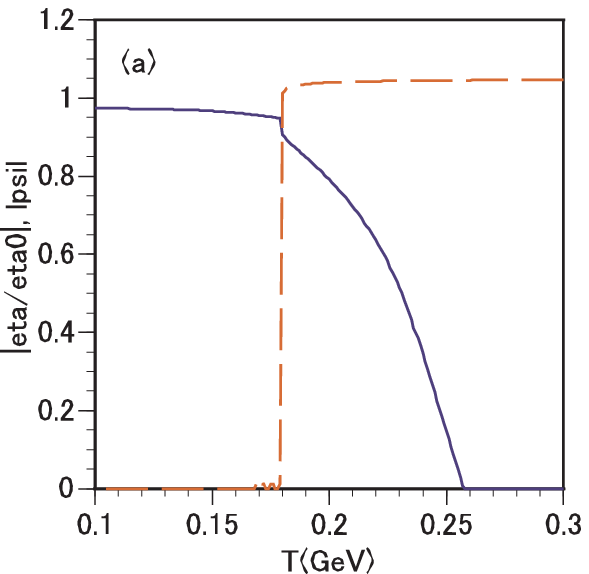} 
 \includegraphics[width=6cm]{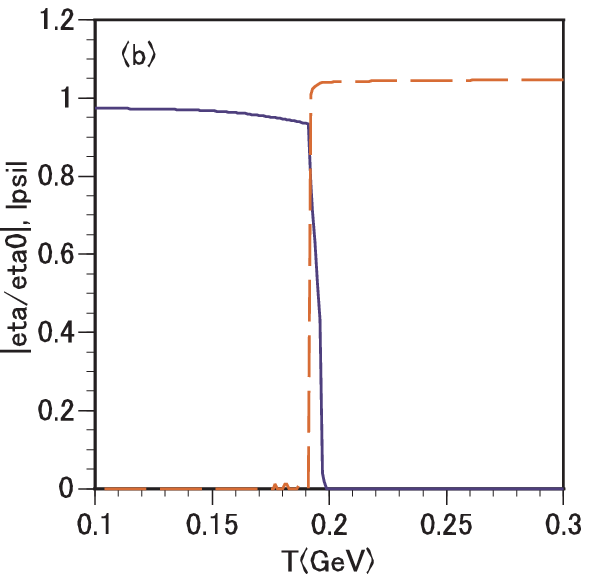} 
\end{center}
\caption{$T$-dependence of the absolute values of the $\eta$ condensate (solid curve) and the phase $\psi$ (dashed curve) of modified Polyakov loop $\Psi$ at $\theta =\pi$ and $\Theta =\pi/3$. 
$\eta$ is normalized by $\eta_0\equiv \eta (T=0)$. 
(a) Original PNJL model with $T_0=200$MeV. (b) EPNJL model with $\alpha_1=\alpha_2=0.2$ and $T_0=200$MeV. 
}
\label{fig2}
\end{figure}

\end{document}